\documentclass{ieeeaccess}
\usepackage{cite} 
\usepackage{amsmath,amssymb,amsfonts}
\usepackage{algorithm}
\usepackage{algorithmic}
\usepackage{graphicx}
\usepackage{textcomp}
\usepackage{amsthm}

\theoremstyle{definition}

\def\BibTeX{{\rm B\kern-.05em{\sc i\kern-.025em b}\kern-.08em
     T\kern-.1667em\lower.7ex\hbox{E}\kern-.125emX}}
\begin{document}


\doi{ ... }

\title{Bound States in Continuum and Zero-Index Metamaterials: A Review}
\author{ 
\uppercase{Yu Peng}\authorrefmark{1, 2},  and 
\uppercase{Shaolin Liao}\authorrefmark{*3, 4}
\IEEEmembership{Senior Member, IEEE}
}
\address[1]{School of Science, Beijing Forestry University, Beijing 100083, China (e-mail: yupeng@g.harvard.edu)}
\address[2]{School of Engineering and Applied Sciences, Harvard University, 9 Oxford Street, Cambridge, Massachusetts 02138, USA}
\address[3]{Argonne National Laboratory, 9700 S. Cass Avenue, Lemont, IL, 60439 USA (e-mail: sliao@anl.gov)}
\address[4]{Department of Electrical and Computer Engineering, Illinois Institute of Technology, 10 West 35th Street, Chicago, IL, 60616 USA (e-mail: sliao5@iit.edu)}
\tfootnote{This work was supported in part by the National Science Foundation of China under Grant No. 11504022, 31530084; Fundamental Research Funds for the Central Universities under Grant No. 2015TP1004.}

\markboth
{Y. Peng \headeretal: Bound States in Continuum and Zero-Index Metamaterials: A Review}
{Y. Peng \headeretal: Bound States in Continuum and Zero-Index Metamaterials: A Review}

\corresp{*Corresponding author: Shaolin Liao (e-mail: sliao@anl.gov).}


\begin{abstract}
Bound states in the continuum (BICs) are waves that remain localized even though
they coexist with a continuous spectrum of radiating waves that can carry energy away.
Their very existence defies conventional wisdom. Although BICs were first proposed in
quantum mechanics, they are a general wave phenomenon and have since been identified
in electromagnetic waves, acoustic waves in air, water waves and elastic waves in solids. These states have been studied in a wide range of material systems, such as piezoelectric materials, dielectric photonic crystals, optical waveguides and fibers, quantum dots, graphene and topological insulators. In this Review, we describe recent developments in this field with an emphasis on the physical mechanisms that lead to BICs across seemingly very different materials and types of waves. We also discuss experimental realizations, existing applications and directions for future work. 
At last, we present our recent effort to design a novel type of silicon (Si) based mu-Near-Zero (uNZ) metamaterials that reduces radiative loss through destructive interference of multiple loss channels, resulting in a bound state in the continuum. This design consists of Si pillars array. By adjusting the unit-cell dimensions including radius, pitch and height, we can eliminate the out-of-plane radiation. 
\end{abstract}


\titlepgskip=-15pt

\maketitle

\section{Introduction}
\label{sec:introduction}
\PARstart{C}{onfinement}, of waves is ubiquitous in nature and in wave-based technology. Examples include electrons bound to atoms and molecules, light confined in optical fibers and the partial confinement of sound in musical instruments. The allowed frequencies of oscillation are known as the wave spectrum. To determine whether a wave can be perfectly confined or not (that is, if a bound state exists or not) in an open system, a simple criterion is to look at its frequency. If the frequency is outside the continuous spectral range spanned by the propagating waves, it can exist as a bound state because there is no pathway for it to radiate away. Conversely, a wave state with the frequency inside the continuous spectrum can only be a resonance that leaks and radiates out to infinity. This is the conventional wisdom described in many books. A bound state in the continuum (BIC) is an exception to this conventional wisdom: it lies inside the continuum and coexists with extended waves, but it remains perfectly confined without any radiation. BICs are found in a wide range of material systems through confinement mechanisms that are fundamentally different from those of conventional bound states.

The general picture is clear from the spectrum and the spatial profile of the modes (Fig. 1). More specifically, consider waves that oscillate in a sinusoidal way as $e^{-i \omega t}$ in time $t$ and at angular frequency $\omega$. Extended states (blue; Fig.1) exist across a continuous range of frequencies. Outside this continuum lie discrete levels of conventional bound states (green; Fig.1) that have no access to radiation channels; this is the case for the bound electrons of an atom (at negative energies), the guided modes of an optical fiber (below the light line) and the defect modes in a bandgap. Inside the continuum, resonances (orange; Fig.1) may be found that locally resemble a bound state but in fact couple to the extended waves and leak out; they can be associated with a complex frequency, $\omega= \omega_0 - i \gamma $, in which the real part $\omega_0$ is the resonance frequency and the imaginary part $\gamma$ represents the leakage rate. This complex frequency is defined rigorously as the eigenvalue of the wave equation with outgoing boundary conditions \cite{Moiseyev},\cite{Kukulin}. In addition to these familiar wave states, there is the less known possibility of BICs (red; Fig.1) that reside inside the continuum but remain perfectly localized with no leakage, namely $\gamma = 0$. In a scattering experiment, waves coming in from infinity can excite the resonances, causing a rapid variation in the phase and amplitude of the scattered waves within a spectral linewidth of $2\gamma$. However, such waves cannot excite BICs, because BICs are completely decoupled from the radiating waves and are invisible in this sense. Therefore, a BIC can be considered as a resonance with zero leakage and zero linewidth ($\gamma = 0$ or infinite quality factor $Q = \omega_0/2\gamma)$. BICs are sometimes referred to as embedded eigenvalues or embedded trapped modes.

\Figure[ht!](topskip=0pt, botskip=0pt, midskip=0pt)[width=0.45\textwidth]{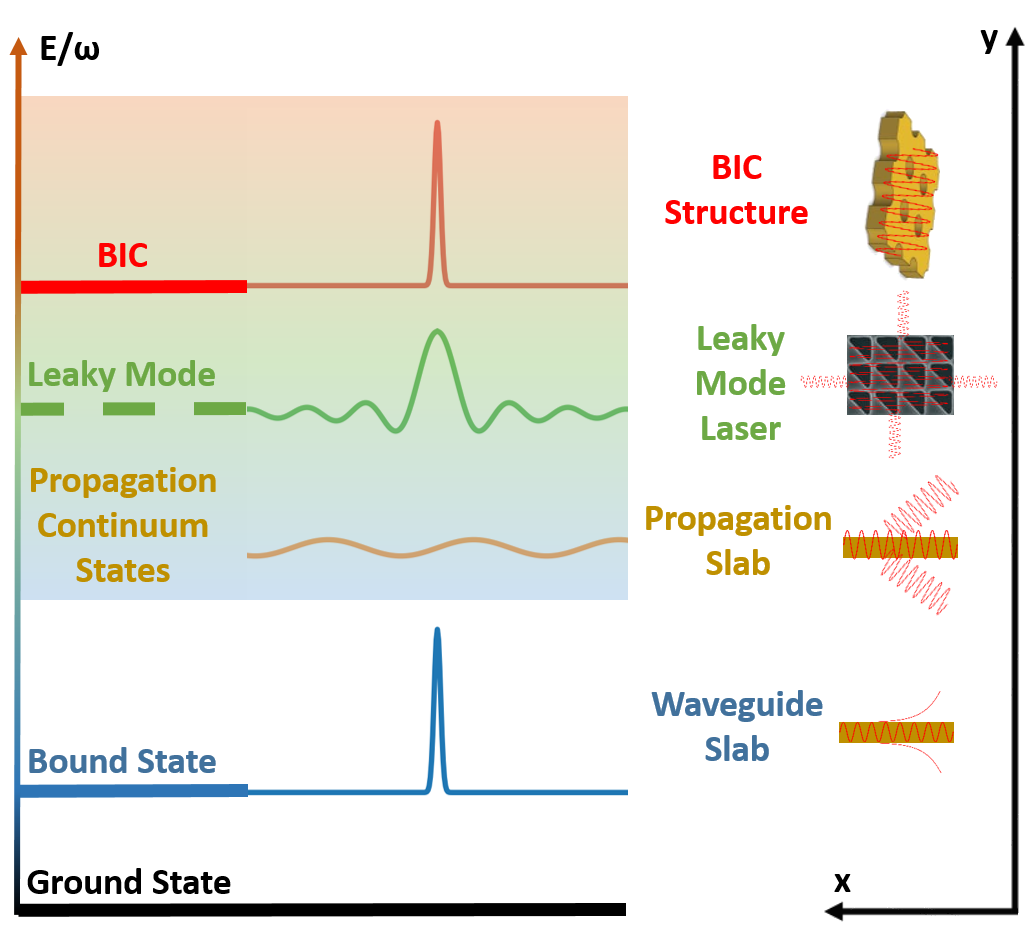}
{ Illustration of different states.\label{fig:BIC}}

In 1929, BICs were proposed by von Neumann and Wigner\cite{Neumann}. As an example, von Neumann and Wigner mathematically constructed a 3D potential extending to infinity and oscillating in a way that was tailored to support an electronic BIC. This type of BIC-supporting system is rather artificial and has never been realized. However, since this initial proposal, other mechanisms leading to BICs have been identified in different material systems, many of which have been observed in experiments in electromagnetic, acoustic and water waves. In recent years, photonic structures have emerged as a particularly attractive platform owing to the ability to tailor the material and structure, which is often impossible in quantum systems. The unique properties of BICs have led to numerous applications, including lasers, sensors, filters and low-loss fibers, with many more possible uses proposed and yet to be implemented. Most theoretically proposed and all experimentally observed BICs are realized in extended structures because, in most wave systems, BICs are forbidden in compact structures. Among the extended structures that support BICs, many are uniform or periodic in one or more directions (for example, x and y), and the BIC is localized only in the other directions (for example, z), as shown in Fig. \ref{fig:BIC}. In such systems, the concept of BICs requires careful definition. More specifically, because translational symmetry conserves the wave vector, $\vec{k} = (k_x, k_y)$, a state is considered a BIC when it exists inside the continuous spectrum of modes at the same $\vec{k}$ but remains localized and does not radiate in the z direction. These BICs are typically found at isolated wave vectors. In this Review, we present the concepts and physical mechanisms that unify BICs across various material systems and in different types of waves, focusing on experimental studies and applications. First, we describe BICs protected by symmetry and separability; second, we discuss BICs achieved through parameter tuning (with coupled resonances or with a single resonance); and third, we describe BICs built with inverse construction (for example, potential, hopping rate or shape engineering). We conclude with the existing and emerging applications of BICs.

\section{BIC Physics}\label{BIC_physics}
BIC is a wave phenomenon that exists in quantum electronics, photonics and acoustics. Without loss of generality, it can be approximated as a scalar Helmholtz problem \cite{Peng_2019,  Liao_2019_IEEE_Sensors, Liao_2020_IEEE_Quantum, Liao_2018_IEEE_Rapid,  Liao_2020_OL, Zeng_2019, Liao_2019, Peng_2019_NEMO} like the Shrodinger equation in quantum mechanism,

\begin{eqnarray}\label{eqn:wave}
- \frac{1}{2} \nabla^2 \psi(\vec{r}) + V(\vec{r}) \psi(\vec{r}) = E \psi(\vec{r}),
\end{eqnarray}
where in photonics system, the potential is given by,
\begin{eqnarray}
V(\vec{r}) = \frac{1}{2} \omega^2 \epsilon \mu - E.
\end{eqnarray}

\subsection{BIC Definition}
BIC mode is defined as the wave confinement when its energy is above the background medium. Confinement here means the wave approaches zero in some specific direction in infinity, while it can propagate along other directions. For example, BIC in $z$ direction means $\psi(z \sim \pm \infty) \sim 0$, when $\omega > c k_0$. So to be a BIC mode, the following conditions have to hold,

\begin{enumerate}
    \item The structure cannot extend to infinity in the BIC confinement direction, i.e., $\exists R \rightarrow  n(|\vec{r}| > R) = n_0$;
    
    \item The energy of the BIC mode has to above the background potential, i.e., $\omega > c k_0$;
    
    \item The wave approaches to zero at infinity in the BIC confinement direction, i.e.,  $\psi(z \sim \pm \infty) \sim 0$ for all real $k_z$.
    
\end{enumerate}

\subsection{BIC Modes and Non-BIC Modes}
Wave confinements that satisfies condition 3) above are called resonant modes. Besides the BIC modes, there are other two types of resonant modes,

\begin{enumerate}

    \item Bound state modes: satisfy condition 1) and violate condition 2) above. Bound state is also called trapped modes because its engy is lower than the background medium. One such example is the dielectric slab waveguide when the waveguide dielectric core wavevector is greater than the background wavevector, i.e., $k > k_0$. In such case, the wave outside the waveguide dielectric core is evanescent and decays to zero in the far field.
    
    \item Photonics Crystal (PhC) modes: satisfy condition 2) and violate condition 1) above. PhC is a periodic structure in the bounded direction where its period $p$ satisfies the Bragg resonance condition, i.e., $p = \lambda/(2 n_{eff})$, with $n_{eff}$ being the effective refractive index. PhC resonance mode is formed when a defect sandwiched between two PhC mirrors.

\end{enumerate}

\subsection{Infinite BIC Structure Potential}
BIC potential can be inversely reconstructed for some specific wave functions, which is usually infinite, 
\begin{align}
\hbox{For the infinitesimal }  & \forall \delta \sim; \  \exists R \in  \hbox{R; such that,} \nonumber \\  \epsilon(|\vec{r}| \ge R ]) &< \delta \sim 0 \rightarrow R \sim \infty. 
\end{align}

Such infinite potential can be readily reconstructed from Eq. (\ref{eqn:wave}) for a given wave function $\psi(\vec{r})$ as follows,
\begin{align}
    V(\vec{r}) = E + \frac{\nabla^2 \psi(\vec{r})}{2 \psi(\vec{r})}.
\end{align}


\subsection{Finite BIC Structure Potential}
It is very difficult to satisfy all above conditions for compact structures. Assuming that the structure has finite boundaries channels can be considered as radiation from equivalent sources $S(\psi(\vec{r}))$ on the structure boundaries $\mathcal{C}$. To satisfy the condition of $\psi(\vec{r} \sim \pm \infty) \sim 0$, the 3-dimension Fourier transform coefficients of $S(\psi(\vec{r}))$ has to be zero. One trivial solution is $S(\psi(\vec{r}))=0$ at every position of the structure boundaries, which contradicts with the continuity requirement imposed by the wave solution of the Helmholtz equation, i.e., the wave solutions inside and outside the structure have to be continuous. 

Now let's look for solutions of compact BIC structures. In terms of wave amplitude and pahse, the wave equation of Eq. (\ref{eqn:wave}) can be re-writen as follows,
\begin{eqnarray}\label{eqn:k}
- \frac{1}{2} \nabla^2 \left(|\psi(\vec{r})| e^{j\phi(\vec{r})} \right) = \left( E - V(\vec{r})  \right) \psi(\vec{r}).  
\end{eqnarray}

Assuming that slowly varying amplitude and locally linear phase, i.e., $\phi(\vec{r}) \sim \vec{k} \cdot \vec{r}= k_x x + k_y y + k_z z$, we Eq. (\ref{eqn:k}) reduces to,
\begin{eqnarray}\label{eqn:k2}
k_x^2(\vec{r}) + k_y^2(\vec{r}) + k_z^2 (\vec{r})   =   k^2(\vec{r}),
\end{eqnarray} 
where $k(\vec{r}) = \sqrt{2 \left(E - V(\vec{r}) \right)}$.

According to Eq. (\ref{eqn:k2}), there are two scenarios for BIC to exist in the bounded direction, e.g., $z$,

\begin{enumerate}
    \item Forbidden propagation BIC: when $k_z$ is imaginary in the background medium, i.e., $k_z = j \gamma$;
    
    \item Wave interference cancellation BIC: when two modes' amplitudes cancel each other in the far field, i.e., $|\psi(|\vec{r}| \sim \infty)| =0$ for all real $k_z$.
\end{enumerate}

\subsection{Forbidden Propagation BIC}
For the forbidden propagation class, the momentum vector at the unbounded directions $k_{//}$ on the structure boundary is greater than the background momentum $k_0$ such that the wave decays exponentially in the bounded state: $\alpha_z = \sqrt{k_{//}^2 - k_0^2}$. One  such forbidden propagation BIC is the symmetry-protected BIC where propagation channels decouple from the bound state due to the symmetry of the structure, as shown in Fig. \ref{fig:symmetry} for the asymmetric BIC acoustic mode: an acoustic plate is inserted within the guided acoustic structure bounded by two acoustic walls. The original acoustic waveguide without the acoustic plate support both even-symmetry modes (Top plot of Fig. \ref{fig:symmetry}) and odd-symmetry modes (Bottom plot of Fig. \ref{fig:symmetry}). For the odd-symmetry modes, when the acoustic energy/frequency is lower than that to support the fundamental oscillation in $y$ direction, i.e., $k < \pi c_s/h$, with $c_s$ being the sound velocity and $h$ the height of the acoustic waveguide, the wavevector along the bounded $x$ direction will become imaginary and the wave is evanescent, i.e., $k_x = j \alpha = j \sqrt{(\pi c_s/h)^2 - k^2}$. Due to symmetry, even-symmetry propagation waveguide propagation modes are orthogonal to the odd-symmetry modes, which causes decoupling and thus BIC odd-symmetric modes, for $k < \pi c_s/h$.

\Figure[ht!](topskip=0pt, botskip=0pt, midskip=0pt)[width=0.45\textwidth]{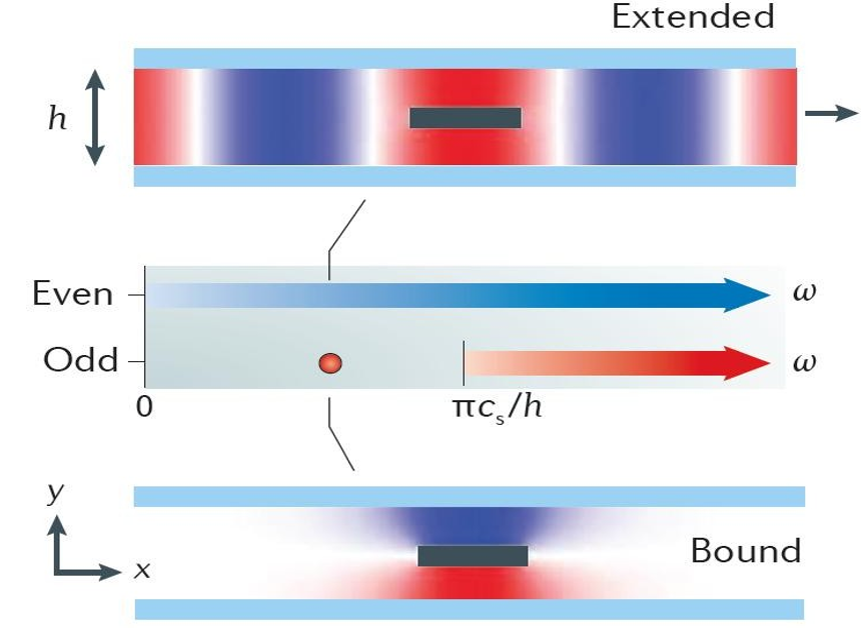}
{ Symmetry-protected BIC.\label{fig:symmetry}}

\subsection{Wave Interference Cancellation BIC}

For the wave interference cancellation class, different leaky modes couple each other to cancel the propagation channel in the far field. To have BIC state, minimum number of  propagation channels is preferred, which can be achieved for periodic structure. If a period  $p < \lambda_0$  along the unbounded direction, only one propagation channel exists. To satisfy the BIC condition of $\psi(z \sim \pm \infty) \sim 0$, at least two Bloch modes are required to cancel each other in the far field.  This class of BIC structures include Fabry-Perot BIC and Friedrich-Wintgen BIC. As an example, let's look at the periodic BIC structure in Fig. \ref{fig:Cancellation}, for $M$ leaky modes and $N$ propagation channels in the unbounded $y$ direction, the wave cancellation in the bounded $x$ direction must satisfy the following,
\begin{eqnarray} 
  \begin{bmatrix}
   \mathcal{F}_{11}   & \mathcal{F}_{12}    & \cdots & \mathcal{F}_{1M}   \\ 
\mathcal{F}_{21}   & \mathcal{F}_{22}    & \cdots & \mathcal{F}_{2M}   \\ 
\vdots  &  \vdots    & \vdots \ \vdots \ \vdots & \vdots   \\ 
\mathcal{F}_{N1}   & \mathcal{F}_{N2}    & \cdots & \mathcal{F}_{NM}   \\ 
  \end{bmatrix}   
  \begin{bmatrix}
  c_1 e^{j \theta_1} \\
    c_2 e^{j \theta_2}  \\
    \vdots  \\
  c_M e^{j \theta_M}
  \end{bmatrix}  = 0  
  \label{eq:BIC_cancel}
\end{eqnarray}
where $\mathcal{F}$ is the Fourier transform coefficients of different leaky modes on the boundary of the periodic BIC structure, for example, $x= 0$ or $x=h$ in Fig. \ref{fig:Cancellation}. $\mathcal{F}$ is usually real for leaky modes with small energy leakage. From Eq. \ref{eq:BIC_cancel}, BIC modes exist only when there are more leaky modes than propagation channels, i.e., $N \ge M$. For the special case of equal number of leaky modes and propagation channles, i.e., $N = M$, one BIC mode exists when and only when the determinant of Fourier transform matrix is zero,
\begin{eqnarray} 
  \begin{vmatrix}
   \mathcal{F}_{11}   & \mathcal{F}_{12}    & \cdots & \mathcal{F}_{1M}   \\ 
\mathcal{F}_{21}   & \mathcal{F}_{22}    & \cdots & \mathcal{F}_{2M}   \\ 
\vdots  &  \vdots    & \vdots \ \vdots \ \vdots & \vdots   \\ 
\mathcal{F}_{N1}   & \mathcal{F}_{N2}    & \cdots & \mathcal{F}_{NM}   \\ 
  \end{vmatrix}    = 0.   
\end{eqnarray}

For single propagation channel, i.e., $N=1$, all leaky modes have the same wavevector $k_x$ in the bounded $x$ direction, we have,
\begin{eqnarray}  
  \sum_{m=1}^M  c_m \mathcal{F}_m e^{j  \theta_m} = 0  .
  \label{eq:BIC_cancel2}
\end{eqnarray}

Without loss of generality, let's look at the case when there are only two leaky modes i.e., $M=2$, as shown in Fig. \ref{fig:BIC}. Eq. (\ref{eq:BIC_cancel2}) reduces to,
\begin{align}
     c_2 \mathcal{F}_2 e^{j  \theta_2} =  -  c_1 \mathcal{F}_1 e^{j  \theta_1},
\end{align}
whose solutions are as follows,
\begin{align}
     c_2    =    c_1 \frac{\mathcal{F}_1}{\mathcal{F}_2}, \ \hbox{and} \ \theta_2 =  \theta_1 + 180^\circ,
\end{align}
from which we can see that the two modes must be out of phase, i.e., the phase difference is 180$^\circ$. It is exactly such out-of-phase interference that causes the cancellation of the BIC modes in far field.

\Figure[ht!](topskip=0pt, botskip=0pt, midskip=0pt)[width=0.45\textwidth]{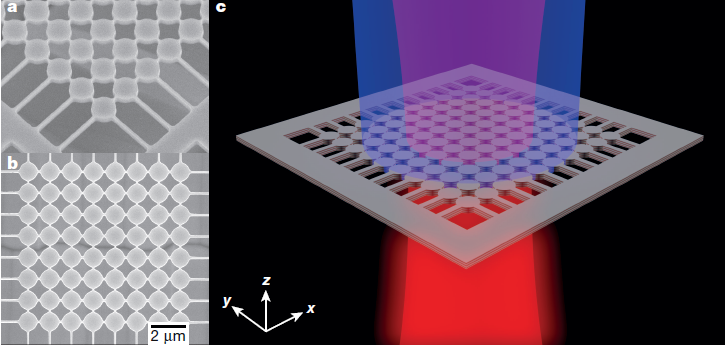}
{The BIC laser: a) Tilted electron micrograph of InGaAsP multiple quantum wells cylindrical nanoresonator array suspended in air. All structures are fabricated using electron beam lithography followed by reactive ion etching to form the cylinders. We subsequently use wet etching to suspend the structure (see section B of Supplementary Information); b) Top view of an 8-by-8 array with supporting bridges, which are used for the mechanical stability of the membrane. The dimensions of the structure are the period (1,200 nm), the thickness (300 nm) and the bridge width (200 nm); and c) Schematic of the fabricated system illustrating the pump beam (blue) and lasing from the BIC mode (red). The radius of the cylindrical nanoresonators is the key parameter in our BIC design.\label{fig:blaser_1}}

  \Figure[ht!](topskip=0pt, botskip=0pt, midskip=0pt)[width=0.45\textwidth]{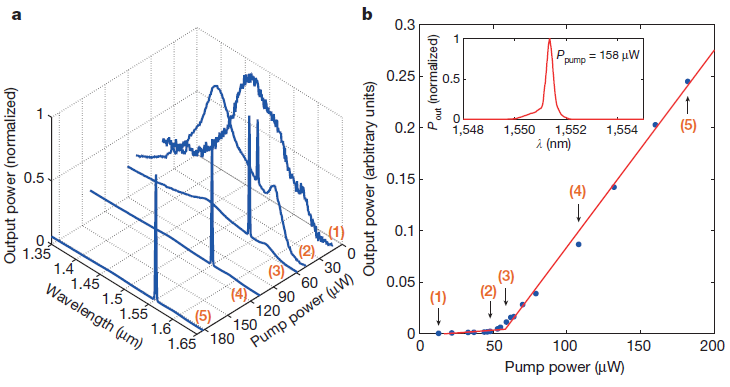}
{Experimental characterization of the
BIC laser: a) Evolution of the normalized output power as a function of wavelength and pump power for a 16-by-16 array with a nanoresonator radius of 525 nm. We observe the transition from a broad spontaneous emission to a single lasing peak at 1551.4 nm; and b) Output power as a function of the average pump power (light-light curve) around the lasing wavelength. We observe the onset of lasing at a threshold power of 56 $\mu$ W. The red lines are linear fits to the data, indicating the regions of spontaneous and stimulated emission. The blue dots correspond to measurements and numbers (1) to (5) denote the spectra plotted in a. Here the standard error in arbitrary units is less than 0.01. The inset shows the lasing spectrum at a pump power of 158 $\mu$W with a measured linewidth of about 0.33 nm (detection-limited).\label{fig:blaser_2}}

\subsection{Lasing action of bound states in continuum}

Ashok Kodigala etc\cite{Ashok_Kodigal}. report, at room temperature, lasing
action from an optically pumped BIC cavity. the results show that the lasing wavelength of the fabricated BIC cavities, each made of an array of cylindrical nanoresonators suspended in air, scales with the radii of the nanoresonators according to the theoretical prediction for the BIC mode, as shown in Fig. \ref{fig:blaser_1}. Moreover, lasing action from the designed BIC cavity persists even after scaling down the array to as few as 8-by-8 nanoresonator, as shown in Fig. \ref{fig:blaser_2}. BIC lasers open up new avenues in the study of
light-matter interaction because they are intrinsically connected
to topological charges\cite{Zhen_B}and represent natural vector beam sources
(that is, there are several possible beam shapes)\cite{Miyai_E}, which are highly
sought after in the fields of optical trapping, biological sensing and
quantum information.

\section{BIC and ZIM}
Let's look at the relation between BIC and ZIM. ZIM requires that $n = \sqrt{\epsilon_r \mu_r} = 0$, which means that,

\begin{enumerate}
    \item eNZ materials: $\epsilon_r \sim 0$.
    \item uNZ materials: $\mu_r \sim 0$. 
    \item eNZ and uNZ materials: $\epsilon_r \sim 0$ and $\mu_r \sim 0$.
\end{enumerate}

\subsection{Effective Refractive Index}
Now let's look at the effective refractive index for composite medium consists of small structures that is small compared with the wavelength. 

\Figure[ht!](topskip=0pt, botskip=0pt, midskip=0pt)[width=0.45\textwidth]{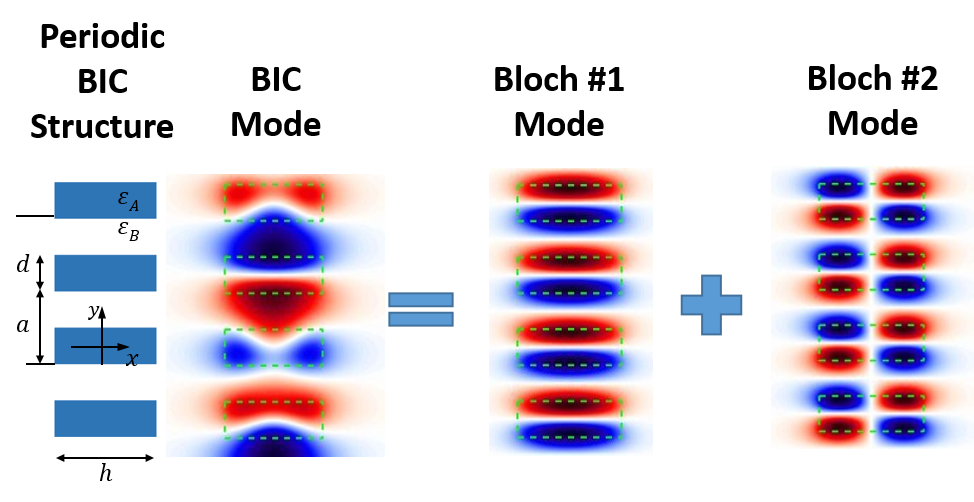}
{ Wave interference cancellation BIC.\label{fig:Cancellation}}

\subsection{Relation between ZIM and BIC}
The wavevector of ZIM materials $\vec{k} \sim 0$, which means $k_x, k_y, k_z \sim 0$. So ZIM mode can be realized through the special type of BIC mode when the unbounded wave vector $k_y =0$.  Under such case, there is only one propagation channel for the BIC mode in Eq. (\ref{eq:BIC_cancel}), which is $k_y = 0$ and $k_x = k_0$ in the background medium. Theoretically, infinite leaky modes will participate in the interference cancellation in this single propagation channel. However, only leaky modes with resonance frequencies close to the BIC resonant frequency have significant contributions. So usually two adjacent leaky modes with resonant frequencies close to the BIC frequency are needed to be considered. ZIM modes are confined in all direction, i.e., $\vec{k} = 0$. At such case, ZIM mode, BIC mode and Bloch mode are the same. Assuming that the Bloch mode's wave function is given by $\psi(x+p, y) =\psi(x, y)$, the radiation in the $y$ direction is given by the Fourier transform of the Bloch wave within the unit cell in $x$ direction, which is $\mathcal{F}(k_x=0, k_y = k_0) = FT_p\{\psi(x, y)\}$. The BIC mode requires that $FT_p\{\psi(x, y)\} =0$, which can be satisfied when the Bloch wave is odd-symmetric within the dielectric element, similar to the symmetry-protected BIC modes. For example, Fig. \ref{fig:BIC_ZIM} shows the dispersion curve (left plot) and Q factor (right plot) of the wave interference cancellation BIC structure given in Fig. \ref{fig:Cancellation}, from which we can see that BIC mode $\textcircled{1}$ and  BIC mode $\textcircled{2}$ are ZIM modes.

BIC modes can also exist even for $\vec{k} \neq 0$. For example, BIC mode $\textcircled{3}$ and BIC mode $\textcircled{4}$ in Fig. \ref{fig:BIC_ZIM} are not ZIM modes.  Theoretically, infinite leaky modes will participate in the interference cancellation in this single propagation channel. However, only leaky modes with resonance frequencies close to the BIC resonant frequency have significant contributions. So usually two adjacent leaky modes with resonant frequencies close to the BIC frequency are needed to be considered. For example,  BIC mode $\textcircled{3}$ in Fig. \ref{fig:BIC_ZIM} can be decomposed into two Bloch modes as in Fig. \ref{fig:Cancellation}.

\Figure[ht!](topskip=0pt, botskip=0pt, midskip=0pt)[width=0.48\textwidth]{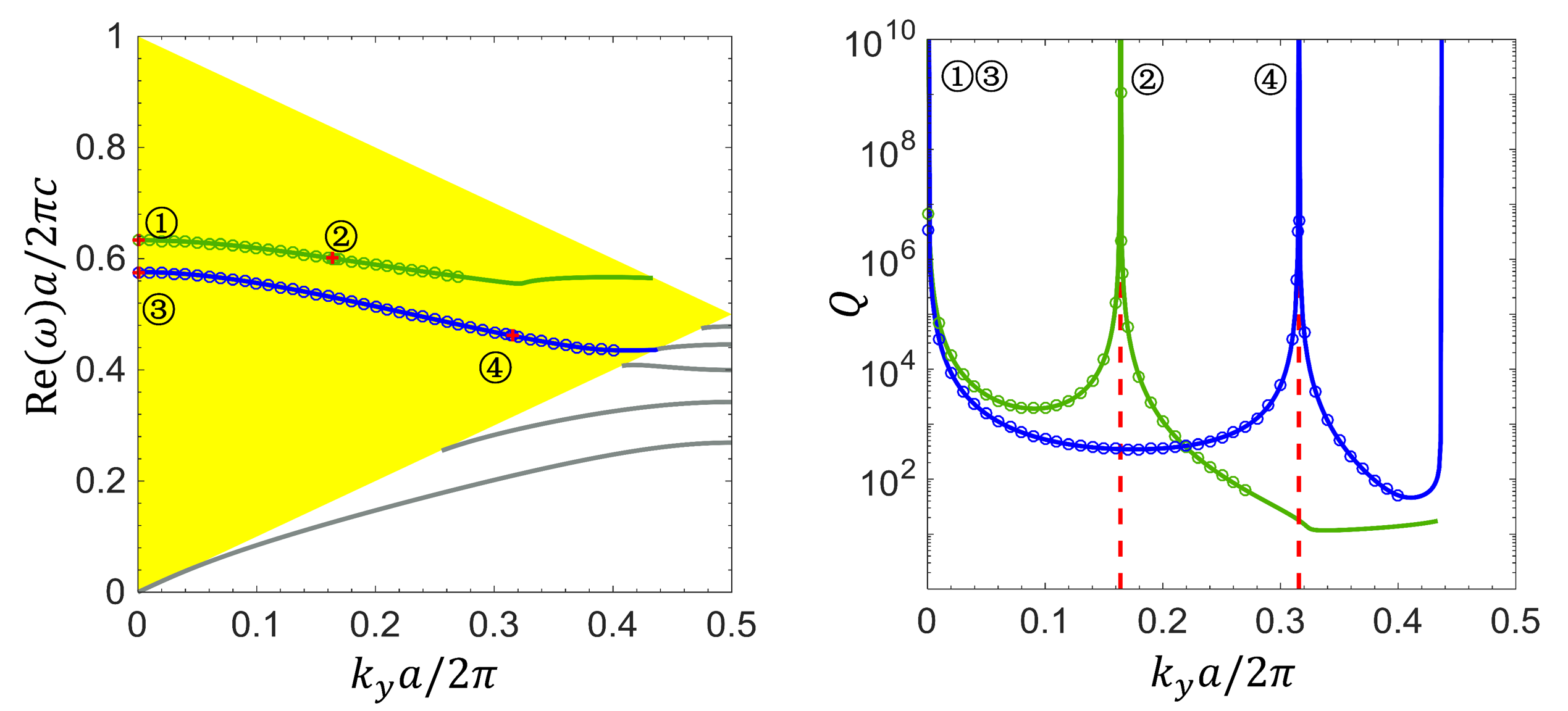}
{Relation between BIC and ZIM.\label{fig:BIC_ZIM}}

\Figure[ht!](topskip=0pt, botskip=0pt, midskip=0pt)[width=0.45\textwidth]{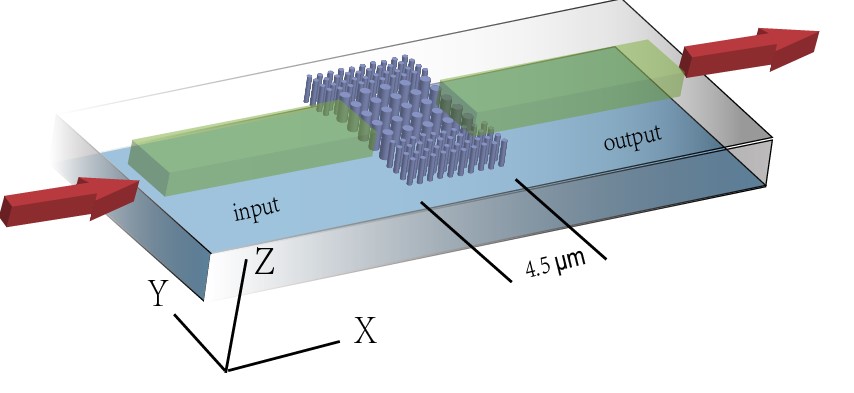}
{The band gap structure of the Si BIC-ZIM. \label{fig:Si_BIC}}

\section{BIC and ZIM on Si Structures}
All zero-index phenomena can be viewed through the lens of the effective wavelength: as the index approaches zero, waves stretch out toward infinity, and ``spatial phase'' looses its meaning as electric and magnetic fields fill the space uniformly. In this sense, the most dramatic applications of zero-index metamaterials will occur in large-area devices that take advantage of the extreme spatial coherence. However, existing demonstrations of integrated
ZIM are only suitable for relatively small length scales, due to absorption during propagation. The absorption stems from ohmic loss in the integrated metal waveguides  which limit the size of metamaterial devices to tens of microns. Access to zero-index
phenomena over macroscopic distances will therefore require a metamaterial design based entirely on lossless dielectrics.
We have recently demonstrated a variation on the original ZIM design that eliminates
the need for metals. This structure consists of a square array of Si pillars on a silica substrate . Removing the metal waveguides causes a frequency shift in the monopole and dipole Mie resonances within the finite pillars, breaking their accidental degeneracy. However, we can restore the degeneracy by adjusting the unit cell dimensions
(radius and pitch) as before, creating an in-plane metamaterial that supports zero-index
modes. Despite the lack of dissipative materials, however, this all-dielectric ZIM suffers
from propagation loss comparable to the metal-based design. Here, the loss is due to out-of-plane radiation from the zero-index mode into free space.
The performance of all-dielectric ZIM highlights a fundamental limitation for any
low-index mode propagating in finite structures, especially modes with zero index. If the
in-plane momentum is less than the momentum of plane waves in free space, then there
exists a continuum mode that is momentum-matched to the guided mode, and light can
couple out of the plane. In the context of the photonic bandstructure these  leaky modes
operate above the light line, and are always associated with radiative loss. This condition
is equivalent to the critical angle for total internal reflection, such that modes above the
light line are inherently unconfined to the substrate. In the case of zero-index modes, it
is impossible to achieve confinement even when the surrounding medium (cladding) has an
arbitrarily low index.
Rather than relying on momentum-based confinement, in this chapter we present a
zero-index metamaterial design that eliminates radiative loss through destructive interference of multiple loss channels, resulting in a bound state in the continuum\cite{D. C. Marinica}\cite{Chia_Nature_2016} . The design
includes an in-plane reflector, creating a Fabry-Perot cavity parallel to the Si pillar array.
By adjusting the coupling between the cavity and the guided mode, we can completely
quench the out-of-plane radiation, resulting in a lossless zero-index metamaterial.

There are several mechanisms beyond total internal reflection that can confine electromagnetic
modes to finite volumes. The simplest is to surround a guiding region with materials
that do not support propagating modes, such a metals or photonic bandgap materials\cite{Attila Mekis}\cite{Joannopoulo1_2008}\cite{Yang Li}.
In this case, there are no continuum modes that can couple to the bound state. Alternatively,
we may design structures whose resonant modes are unable to couple to continuum modes
due to symmetry mismatch, such as dark state lasers\cite{Gentry_2014}\cite{Aditya Jain} . Finally, bound states can occur
in structures which couple to multiple continuum modes, as long as the various loss channels
interfere destructively\cite{Yi Yang}. Such bound states in the continuum (BICs) were first observed in
quantum mechanical systems\cite{Neumann}, and have been extended to integrated photonics \cite{D. C. Marinica}\cite{Yonatan Plotnik}\cite{ Thomas Lepetit}\cite{Chia_2013}\cite{Chia_Nature_2013}.
Because the confinement does not rely on any specific modal symmetry, photonic BICs can
arise in Bloch modes operating above the light line.
In order to create a BIC in dielectric ZIM, we must identify the sources of radiative
loss and arrange for them to interfere. The zero-index mode is the result of accidental
degeneracy of monopole and dipole Mie resonances at the center of the Brillouin zone. The
symmetry of the monopole mode prevents coupling to the continuum; however, the dipole
mode is compatible with plane waves, which radiate from both sides of the pillar array. Therefore we can achieve a BIC by adding a reflector above the plane of the
array, which couples the two loss channels. Energy that escapes through the top surface is
reflected back toward the pillar array, where it couples back into the guided mode or into
the space below.

Bound states above the continuum threshold has been shown by von Neumann and Wigner through the Schrodinger's equation \cite{Maas_2013}. These peculiar states, known as the bound states in the continuum (BIC), exhibit a resonance that does not decay. Decades later, the idea was dormant, primarily as a mathematical curiosity. In 1977, when Herrick and Stillinger suggested that the BIC be observed in the semiconductor superlattice, they were re-interested in the BIC\cite{Maas_2013}. On the other hand, nanostructured composites including metamaterials are used to achieve new refractive phenomena by adjusting the effective refractive index of the medium. In extreme cases, the index can be close to zero, resulting in waves propagating at infinite phase velocities and infinite wavelengths \cite{Maas_2013}. When integrated with on-chip photonic circuits, zero refractive index provides unprecedented phase control in the field of nonlinear and quantum optics\cite{Maas_2013}. Recently, the metamaterials combine zero index property at the center of the Brillouin zone ($\Gamma$-point) and bound states in the continuum (BIC) at $\Gamma$-point, which quench radiative loss at $\Gamma$-point around zero refractive index\cite{Maas_2013}.

	By adjusting the height of the Si pillar array, we can completely quench the out-of-plane radiation, resulting in a lossless zero-index metamaterial.
In order to design a zero-index metamaterial, we adopt an effective medium theory based on scattering cancellation in Mie resonant structures. Briefly, the array of cylinders is treated as an effective medium with permittivity and permeability and , respectively. We consider a single dielectric cylinder surrounded by an annulus of background dielectric \cite{Maas_2013}. We can solve for the Mie scattering coefficients corresponding to the monopole and dipole modes under the assumption that the operating wavelength in the effective medium is much larger than the size of the cylinder and the annular coating. The effective medium approximation is valid when both of these scattering coefficients vanish, in which case an infinite array of cylinders can support propagating Bloch waves. By enforcing this no-scattering condition, we arrive at the effective medium parameters \cite{Maas_2013},
where is the ratio of incident and scattered fields in the background medium for the mth Mie resonance,
with   being the wavevector in the background dielectric,  being the radius of the background dielectric, and  being the radius of the cylinder. The refractive index within the cylinder and the background are   and  , respectively.   is the Bessel function of the second kind. For a given array of dielectric cylinders, Eq. (\ref{eqn:e2}) and Eq. (\ref{eqn:e3})  give the frequency-dependent effective permittivity and permeability. Further, we may invert this process to determine the metamaterial structure that achieves an effective index equal to zero. Simplifying equations in the limit  $\epsilon_{eff} \rightarrow 0$ and   $\mu_{eff} \rightarrow 0$  results in a pair of conditions that must be satisfied in order to achieve $n = 0$.
  	We note that the left-hand side of each condition is a function of the size of the background region relative to the operating wavelength  . Similarly, the right-hand side depends only on the size of the cylinder relative to the operating wavelength , as well as the refractive index of the cylinder . Therefore, these two relations determine a family of solutions for the two design parameters   and   that will generate an effective index equal to zero, which can be satisfied for cylinders with sufficiently high index . High-index inclusions are necessary to ensure that the dipole resonance is excited at the operating wavelength, without which there can be no magnetic response. The residual is minimized when the effective permittivity and permeability are simultaneously zero. In this case, the magnitude of the effective index is also zero.
When light passes through the medium, some part of it will always decay. This can be easily taken into account by defining the complex refractive index,  . Here, the real part   is the refractive index and represents the phase velocity, while the imaginary part   is called the extinction coefficient. Although   can also refer to the mass attenuation coefficient and indicate the attenuation amount when the electromagnetic wave travels through the material. The permittivity can be expressed as  , where the real part is  and  .
 	By tuning the height of the pillar array, we can achieve perfect confinement of the dipole mode. However, the height of pillar array also affects the resonant frequency, as predicted by temporal coupled mode theory (tCMT)\cite{Maas_2013}. The height simultaneously affects the confinement and frequency of the dipole mode. The monopole mode, which does not couple with the continuum, maintains a constant resonant frequency. In order to support zero-index propagation, the monopole and dipole modes must be degenerate, which can only be achieved for a particular height of pillar array. Lossless zero-index propagation requires both degeneracy and confinement for the same height. We may achieve accidental degeneracy of both conditions by adjusting the geometry of the pillar array (e.g. radius, pitch). The lossless ZIM design consists of a square array of 1263-nm-tall Si pillars, with a radius r = 183.3 nm and pitch a = 924.8 nm. 
 	
In this paper we explore bound state in the continuum with zero refractive index in more general cases rather than a specific point Dirac cone. With designing uNZ, we eliminate radiative losses through destructive interference of multiple loss channels, resulting in a bound state in the continuum. The design consists of a Si pillar array, as shown in Fig. \ref{fig:Si_BIC}. By adjusting the unit-cell dimensions including radius, pitch and height, we can completely eliminate the out-of-plane radiation.

\section{Theoretical Formulation of the Si eNZ/uNZ}
The electromagnetic wave equation  describes the full richness of optics - how light bounces and bends while propagating through space-yet the interaction of light and matter is parameterized by just two quantities: one is the electric permittivity \(\varepsilon \) , which describes the way that materials respond to electric fields. The other is its dual, the magnetic permeability \(\mu \), which describes the response to magnetic fields. Together these constitutive parameters describe phenomena ranging from refraction to plasmonics. Even the refractive index \(n=\sqrt{\varepsilon \mu }\) , which measures the wave speed in a material, relies on the complex interplay of electric and magnetic effects.
\par
\begin{eqnarray}\label{eqn:e1}
{{\Delta }^{2}}E-\frac{\mu \varepsilon }{{{c}^{2}}}\frac{{{\partial }^{2}}E}{\partial {{t}^{2}}}=0.
\end{eqnarray}

In this sense, one can map the entire landscape of optical materials and phenomena along two dimensions, and perhaps uncover new physics in unexplored regions of this parameter space.
The origin of this unusual behavior is a structural change in the underlying physics. Optical materials are generally governed by a wave equation, which relates temporal and spatial gradients of the electric and magnetic fields by a proportionality constant equal to the refractive index. When the index approaches zero, the spatial and temporal behavior become decoupled. The spatial field distribution is no longer described by a wave equation, but by Laplace's equation, \({{\Delta }^{2}}E=0\) . This is the physics that governs electrostatics, but crucially, it becomes accessible at optical frequencies. The result is a material that behaves like an electronic conductor, but which channels light instead of current.
Coupling efficiency into a material is determined by the impedance mismatch across the interface \cite{Born}. The reflection vanishes if the impedance is the same for both materials, and it gets worse as the mismatch grows. In general, the impedance  \(\eta =\sqrt{\frac{\mu }{\varepsilon }}\) depends on the ratio of the permittivity and permeability.  When \(\varepsilon =0\)and \(\mu \ne 0\) , the impedance becomes enormous and the reflectivity of the surface approaches 100 percent. When \(\mu =0\) , the impedance becomes small and the reflectivity of the surface is small, and the impedance maintains a finite value.
\par

     In order to design a zero-index metamaterial, we adopt an effective medium theory based on scattering cancellation in Mie resonant structures. Briefly, the array of cylinders is treated as an effective medium with permittivity and permeability \({{\varepsilon }_{\text{e}ff}}\)and \({{\mu }_{\text{e}ff}}\) , respectively. We consider a single dielectric cylinder surrounded by an annulus of background dielectric \cite{Sakoda_OE_2012} \cite{Sakoda_OE_2011} . We can solve for the Mie scattering coefficients corresponding to the monopole and dipole modes under the assumption that the operating wavelength in the effective medium is much larger than the size of the cylinder and the annular coating. The effective medium approximation is valid when both of these scattering coefficients vanish, in which case an infinite array of cylinders can support propagating Bloch waves. By enforcing this no-scattering condition, we arrive at the effective medium parameters \cite{Wu}: 
     
\begin{eqnarray}\label{eqn:e2}
\frac{{{\varepsilon }_{\text{e}ff}}+2\frac{{{{{J}'}}_{0}}({{k}_{0}}{{r}_{0}})}{{{k}_{0}}{{r}_{0}}{{J}_{0}}({{k}_{0}}{{r}_{0}})}}{{{\varepsilon }_{\text{e}ff}}+2\frac{{{{{Y}'}}_{0}}({{k}_{0}}{{r}_{0}})}{{{k}_{0}}{{r}_{0}}{{Y}_{0}}({{k}_{0}}{{r}_{0}})}}=\frac{{{Y}_{0}}({{k}_{0}}{{r}_{0}})}{i{{J}_{0}}({{k}_{0}}{{r}_{0}})} \left(\frac{{{D}_{0}}(\omega )}{1+{{D}_{0}}(\omega )} \right),
\end{eqnarray}  
          
\begin{eqnarray}\label{eqn:e3}
\frac{{{\mu }_{\text{e}ff}}+2\frac{{{J}_{1}}({{k}_{0}}{{r}_{0}})}{{{k}_{0}}{{r}_{0}}{{{{J}'}}_{1}}({{k}_{0}}{{r}_{0}})}}{{{\mu }_{\text{e}ff}}+2\frac{{{Y}_{1}}({{k}_{0}}{{r}_{0}})}{{{k}_{0}}{{r}_{0}}{{{{Y}'}}_{1}}({{k}_{0}}{{r}_{0}})}}=\frac{{{{{Y}'}}_{1}}({{k}_{0}}{{r}_{0}})}{i{{{{J}'}}_{1}}({{k}_{0}}{{r}_{0}})} \left(\frac{{{D}_{1}}(\omega )}{1+{{D}_{1}}(\omega )} \right).
\end{eqnarray}
where \({{D}_{m}}(\omega )\)is the ratio of incident and scattered fields in the background medium for the $m$-th Mie resonance:
\begin{flalign}\label{eqn:e4} 
  & {{D}_{m}}(\omega )=\frac{{{b}_{m}}}{{{a}_{m}}} \\ 
 & =\frac{{{n}_{c}}{{{{J}'}}_{m}}({{n}_{c}}{{k}_{0}}{{r}_{c}}){{J}_{m}}({{k}_{0}}{{r}_{c}})-{{J}_{m}}({{n}_{c}}{{k}_{0}}{{r}_{c}}){{{{J}'}}_{m}}({{k}_{0}}{{r}_{c}})}{{{J}_{m}}({{n}_{c}}{{k}_{0}}{{r}_{c}})H_{m}^{(1)}({{k}_{0}}{{r}_{c}})-{{n}_{c}}{{{{J}'}}_{m}}({{n}_{c}}{{k}_{0}}{{r}_{c}})H_{m}^{(1)}({{k}_{0}}{{r}_{c}})}. \nonumber 
\end{flalign}
with \({{k}_{0}}\) being the wavevector in the background dielectric, \({{r}_{0}}\) being the radius of the background dielectric, and \({{r}_{c}}\) being the radius of the cylinder. The refractive index within the cylinder and the background are  \({{n}_{c}}\)and  \({{n}_{0}}\), respectively. \({{Y}_{m}}\) is the Bessel function of the second kind. For a given array of dielectric cylinders,    and  \label{def:e2} give the frequency-dependent effective permittivity and permeability. Further, we may invert this process to determine the metamaterial structure that achieves an effective index equal to zero. Simplifying equations in the limit ${{\varepsilon }_{\text{e}ff}} \rightarrow 0$ and ${{\mu }_{\text{e}ff}} \rightarrow 0$ results in a pair of conditions that must be satisfied in order to achieve $n = 0$,
                                                    	    
\begin{align}\label{eqn:e5}
\frac{i{{{{J}'}}_{0}}({{k}_{0}}{{r}_{0}})}{{{{{Y}'}}_{0}}({{k}_{0}}{{r}_{0}})}=\frac{{{D}_{0}}({{k}_{0}}{{r}_{c}})}{1+{{D}_{0}}({{k}_{0}}{{r}_{c}})}\end{align}                                                     	      \begin{align}
\frac{i{{J}_{1}}({{k}_{0}}{{r}_{0}})}{{{Y}_{1}}({{k}_{0}}{{r}_{0}})}=\frac{{{D}_{1}}({{k}_{0}}{{r}_{c}})}{1+{{D}_{1}}({{k}_{0}}{{r}_{c}})}.
\end{align}

\par
We note that the left-hand side of each condition is a function of the size of the background region relative to the operating wavelength, $k_0 r_0$. Similarly, the right-hand side depends only on the size of the cylinder relative to the operating wavelength, $k_0 r_c$, as well as the refractive index of the cylinder $n_c$ . Therefore, these two relations determine a family of solutions for the two design parameters  \({{k}_{0}}{{r}_{c}}\) and \({{k}_{0}}{{r}_{c}}\)  that will generate an effective index equal to zero, which can be satisfied for cylinders with sufficiently high index \({{n}_{c}}\). High-index inclusions are necessary to ensure that the dipole resonance is excited at the operating wavelength, without which there can be no magnetic response. The residual is minimized when the effective permittivity and permeability are simultaneously zero. In this case, the magnitude of the effective index is also zero.

Optical metamaterials-composite materials made of nanoscale building blocks can achieve an unconventional effective index of refraction by engineering the geometry of the building blocks. In extreme cases, the effective index can approach zero, resulting in infinite phase velocity and spatial wavelength \cite{Huang} \cite{Moitra}. However, because the zero index corresponds to an electric monopole mode and two magnetic dipole modes degenerating at the center of the Brillouin zone, it opens a loss channel to the direction perpendicular to the substrate, i.e. the guided wave propagating within the zero-index metamaterial can couple to a plane-wave radiating in the out-of-plane direction (the direction perpendicular to the substrate). To handle this, some works have been reported through about eliminating this out-of-plane radiation loss via the destructive interference between the plane waves radiating upward and downward, forming a bound state in the continuum \cite{Kita}\cite{Hsu}. A zero-index metamaterial are designed with a reflector that reduces radiation loss through destructive interference from multiple lossy channels, resulting in continuous bound states at Dirac cone \cite{Philip}\cite{Philip2}. For zero-index metamaterials, there are two types. One is called epsilon-near-zero metamaterials (eNZ) which means electric permittivity equaling to zero. Another is called mu-near-zero metamaterials (uNZ), which means magnetic permeability equaling to zero. 

We use COMSOL Multiphysics software to scan the band structure and get band structure around the Brillouin zone and tune the Si parameters to obtain the  Dirac cone at the center of the Brillouin region.

\Figure[ht!](topskip=0pt, botskip=0pt, midskip=0pt)[width=0.45\textwidth]{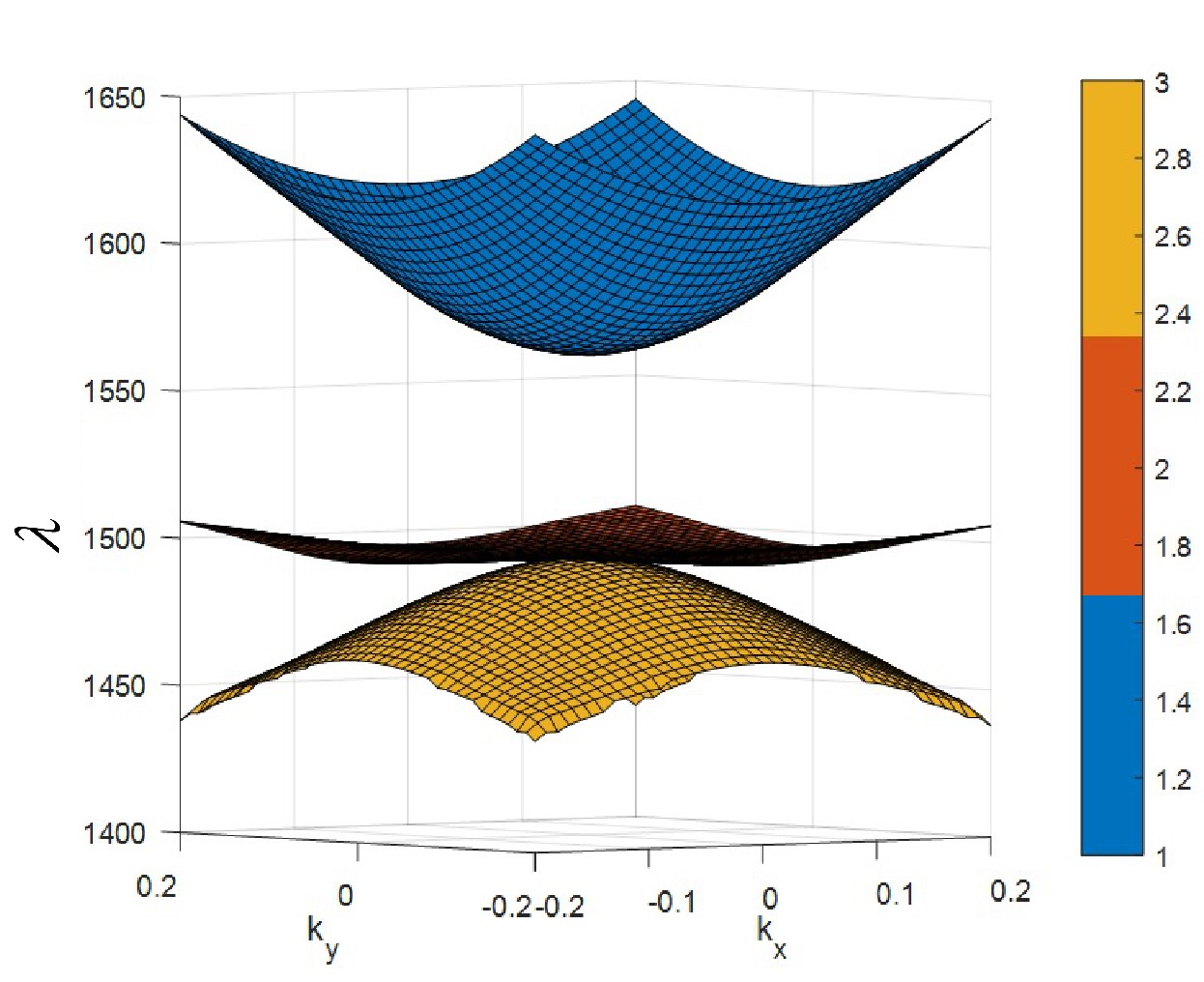}
{ The conventional photonic bands at the center of the Brillouin zone. The blue band represents the propagating band I; the red band represents the propagating band II; the yellow band represents the propagating band III separately.\label{fig:bands_1}}

\Figure[ht!](topskip=0pt, botskip=0pt, midskip=0pt)[width=0.45\textwidth]{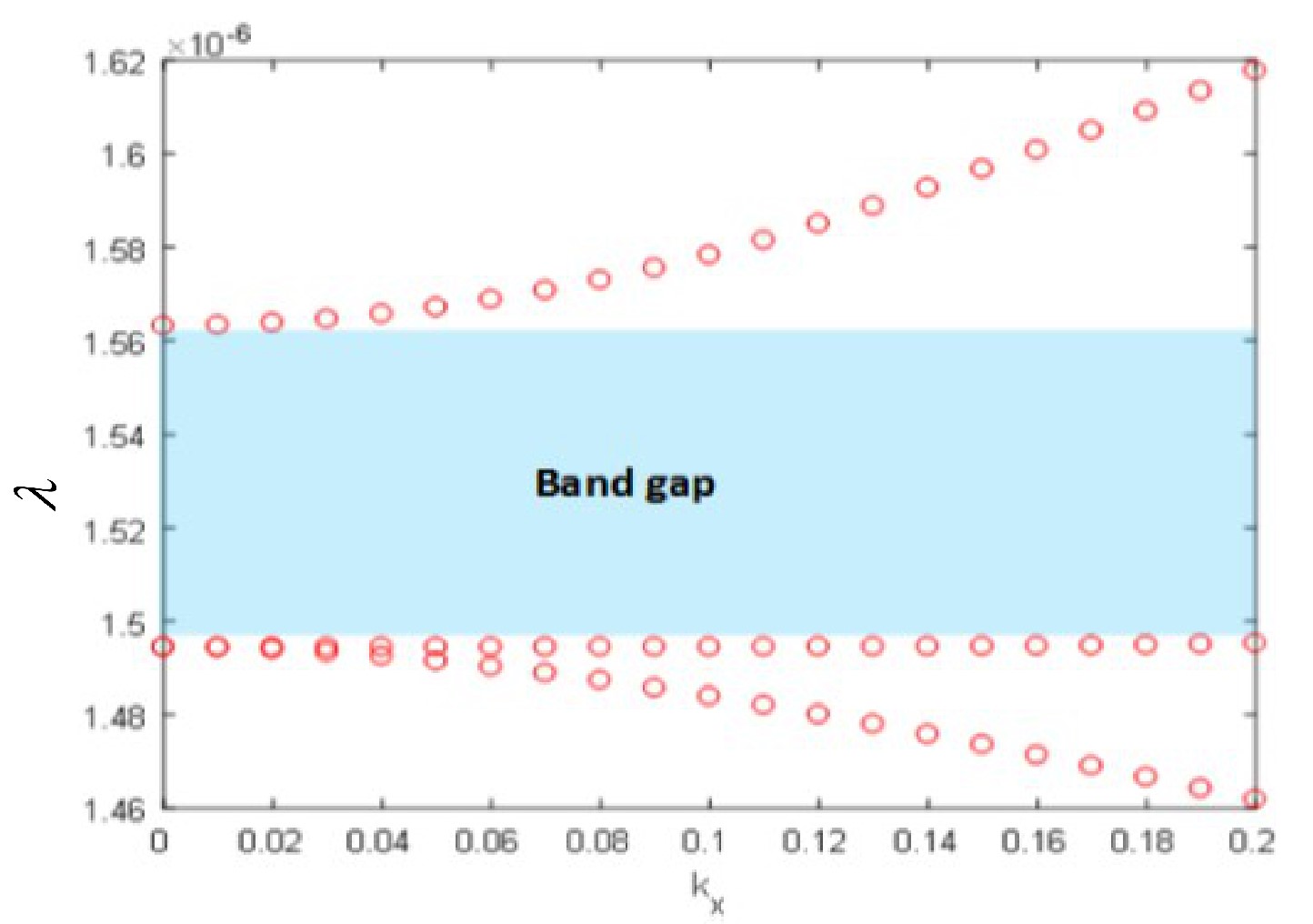}
{ The conventional photonic bands around the center of the Brillouin zone.  (c) Mode profiles at the $\Gamma$-point, corresponding to a monopole (M, band I) and two dipoles (Dx band II and Dy band III). The colormap shows the out-of-plane electric field on a cross-section through the unit cell.\label{fig:bands_2}}

\section{The Conventional PhC Bandgap Structure}
Without tuning of the Si parameters, the Si pillar array shows the conventional PhC bandgap structure.  Fig. \ref{fig:bands_1} and Fig. \ref{fig:bands_2} show the conventional PhC band structure of the presented metamaterial. Near the center of the Brillouin zone, the three bands correspond to a monopole mode and two orthogonal dipole modes, respectively. The bandgap clearly shows up, which is typical for a PhC structure.

\section{The BIC uNZ metamaterials}\label{uNZ} 
The uNZ must include contributions from two dipole modes with mutually orthogonal polarizations. Each dipole couples to continuum modes (vertically propagating plane waves) whose electric field is polarized perpendicular to the magnetic moment of the dipole mode. Thus, each dipole accounts for two of four possible radiative channels, corresponding to two orthogonal polarizations above \cite{Nagai} \cite{Wu2}  and below the pillar array. Since uNZ design retains the two-fold rotational symmetry of the pillar array, the condition is satisfied for both dipole modes automatically. Therefore, all four loss channels are eliminated. Furthermore, Bloch modes near the $\Gamma$-point can always be represented as a linear combination of the two orthogonal dipoles regardless of the propagation direction \cite{Xueqin} \cite{Sakoda_OE_2012} \cite{Simovski_2007} \cite{Engheta1_Science_2003} \cite{Suchowski} \cite{Gellineau_OE_2014}  \cite{Salvatore_Savo}.     

Under the excitation of in-plane transverse magnetic mode (TM), the pillars support three Bloch modes in the center of Brillouin region: Axial electric monopole mode and two transverse magnetic dipole modes. By adjusting the radius and spacing of the array, we can achieve uNZ in the center of the Brillouin region and at the given working wavelength, in our example 1495nm  \cite{Huang}\cite{Moitra} \cite{Kita}. 

We begin this design with a square array of Si pillars having radius r= 183 nm, pitch a= 924.8 nm, and height h = 1263 nm. To maximize Q factor of a uNZ metamaterials, we optimize the pitch of the array and height of the pillars. The final design is with r=170 nm, a=924.8 nm, and height h=1263 nm. Ideally the array should be designed to be 20*20 pillars array to demonstrate the results. But in terms of deceasing the simulation data and for simplification, we simulate 5*5 pillars array to get the phenomenon rather than 20*20 pillars array with the symmetric boundary condition at Y direction. 

\Figure[ht!](topskip=0pt, botskip=0pt, midskip=0pt)[width=0.45\textwidth]{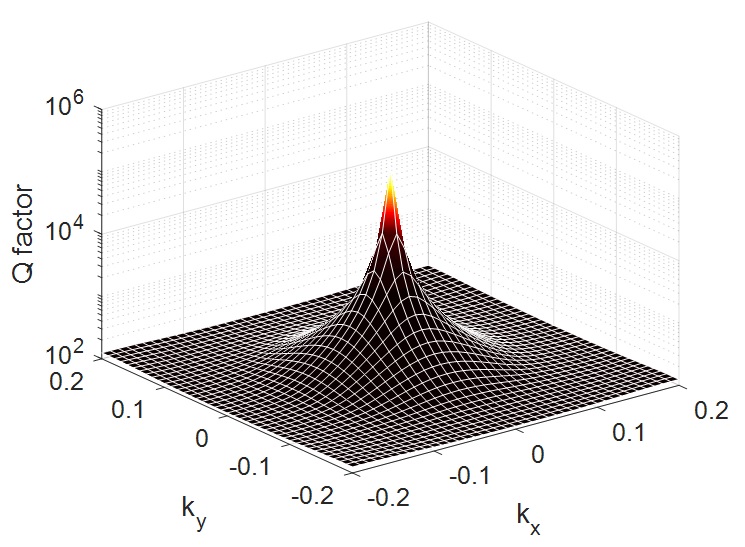} 
{ The color represents the quality factor of the propagating mode 1, where the highest value is up to 105.\label{fig:Q_mode_1}}

 \Figure[ht!](topskip=0pt, botskip=0pt, midskip=0pt)[width=0.45\textwidth]{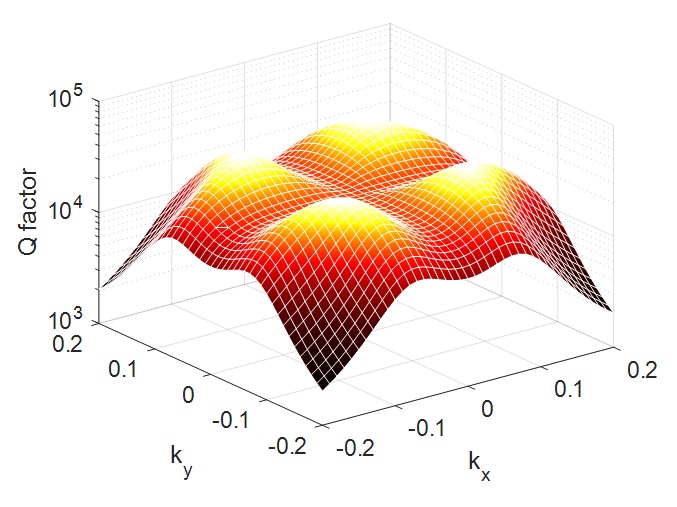}
{ The color represents the quality factor of the propagating mode 2, where the highest value is up to 104.\label{fig:Q_mode_2}}

 \Figure[ht!](topskip=0pt, botskip=0pt, midskip=0pt)[width=0.45\textwidth]{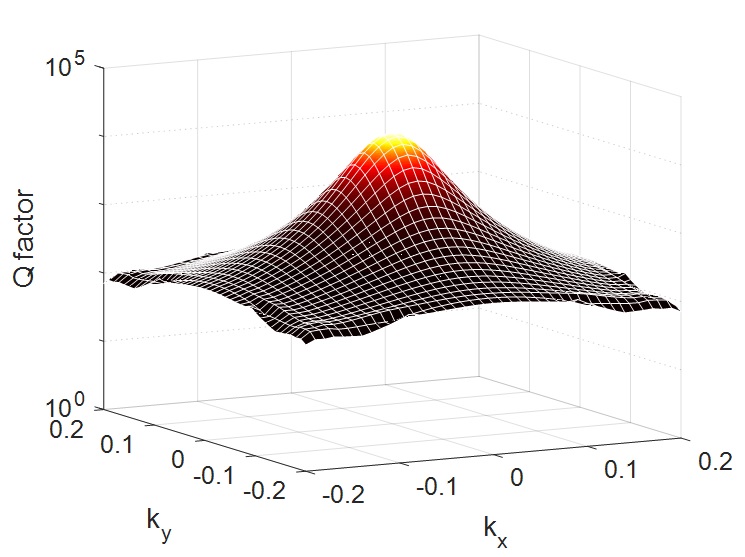}
{The color represents the quality factor of the propagating mode 3, which is strongly peaked at  $\Gamma$-point.\label{fig:Q_mode_3}}

After the uNZ parameters optimization, the isotropic high Q arises due to the rotational symmetry of the lattice, and is specific to the $\Gamma$-point, as shown in Fig. \ref{fig:Q_mode_1}. This design shows high-Q zero index around 1563 nm, near which only the electric monopole mode appears. The monopole mode is by nature high Q because it cannot couple to the out-of-plane radiation. This electric monopole mode corresponds to electric permittivity equaling to zero.
     contours indicate that the behavior is isotropic near the wavelength of design. The quality factor has a strong peak near this wavelength and is isotropic. Optimized design achieved Q > $10^5$,
, up to two orders of magnitude higher than previous designs \cite{Philip2}

\par

The dipole modes become lossless simultaneously at an operating wavelength of 1495 nm. The optimized design achieves Q > $10^4$ , as a result, the optimized metamaterial supports a zero-index mode at 1495 nm, which propagates without phase advance or radiative loss.  Fig. (\ref{fig:Q_mode_2}), and Fig. (\ref{fig:Q_mode_3}) indicate the Q factor at each point in k-space. The overlap at the operating wavelength indicates that the modes are simultaneously triggered and lossless (\ref{fig:bands_1}) \cite{Nagai}.   The zero-index mode must include contributions from two dipole modes with mutually orthogonal polarizations. Each dipole couples to continuum modes (vertically propagating plane waves) whose electric field is polarized perpendicular to the magnetic moment of the dipole mode. Thus, each dipole accounts for two of four possible radiative channels, corresponding to two orthogonal polarizations above and below the pillar array. Since the BIC uNZ design retains the two-fold rotational symmetry of the pillar array, the BIC condition is satisfied for both dipole modes automatically. Therefore, all four loss channels are eliminated, as shown in Fig. \ref{fig:field_xy} and Fig. \ref{fig:field_xz}.

 \Figure[ht!](topskip=0pt, botskip=0pt, midskip=0pt)[width=0.45\textwidth]{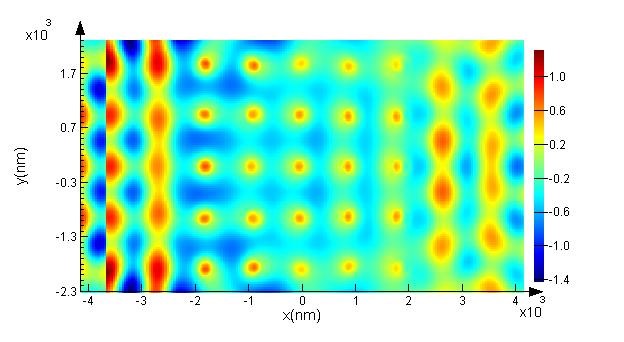}
{ BIC uNZ inplane profile. The field profile shows the vertical component of the electric field through a horizontal cross-section of the array at 1495 nm.\label{fig:field_xy}}

 \Figure[ht!](topskip=0pt, botskip=0pt, midskip=0pt)[width=0.45\textwidth]{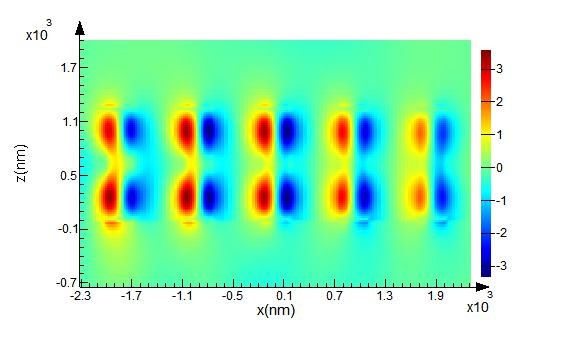}
{ BIC uNZ out-plane profile. The field profile shows the vertical component of the electric field through a vertical cross-section of the array at 1495 nm.\label{fig:field_xz}}

\par

 \Figure[ht!](topskip=0pt, botskip=0pt, midskip=0pt)[width=0.45\textwidth]{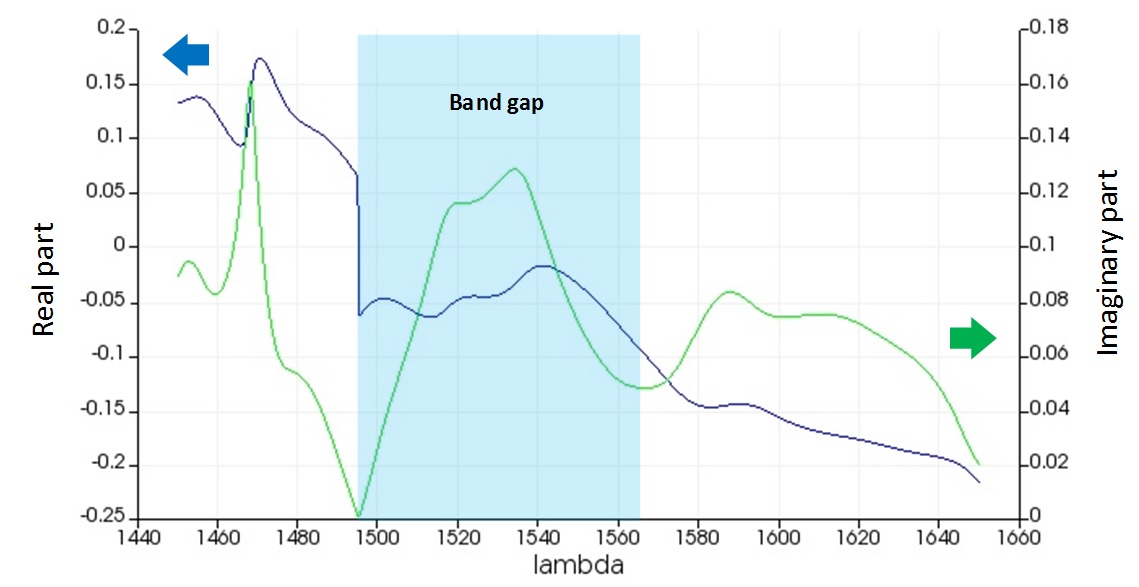}
{ The value of the refractive index of BIC uNZ. Green line is imaginary part, blue line is real part.\label{fig:refractive_index}}

 \Figure[ht!](topskip=0pt, botskip=0pt, midskip=0pt)[width=0.45\textwidth]{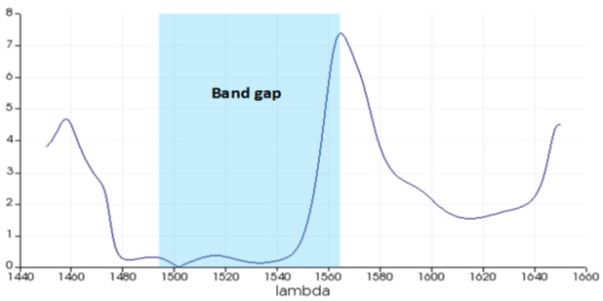}
{The impedance versus wavelength.\label{fig:impedance}}

 \Figure[ht!](topskip=0pt, botskip=0pt, midskip=0pt)[width=0.45\textwidth]{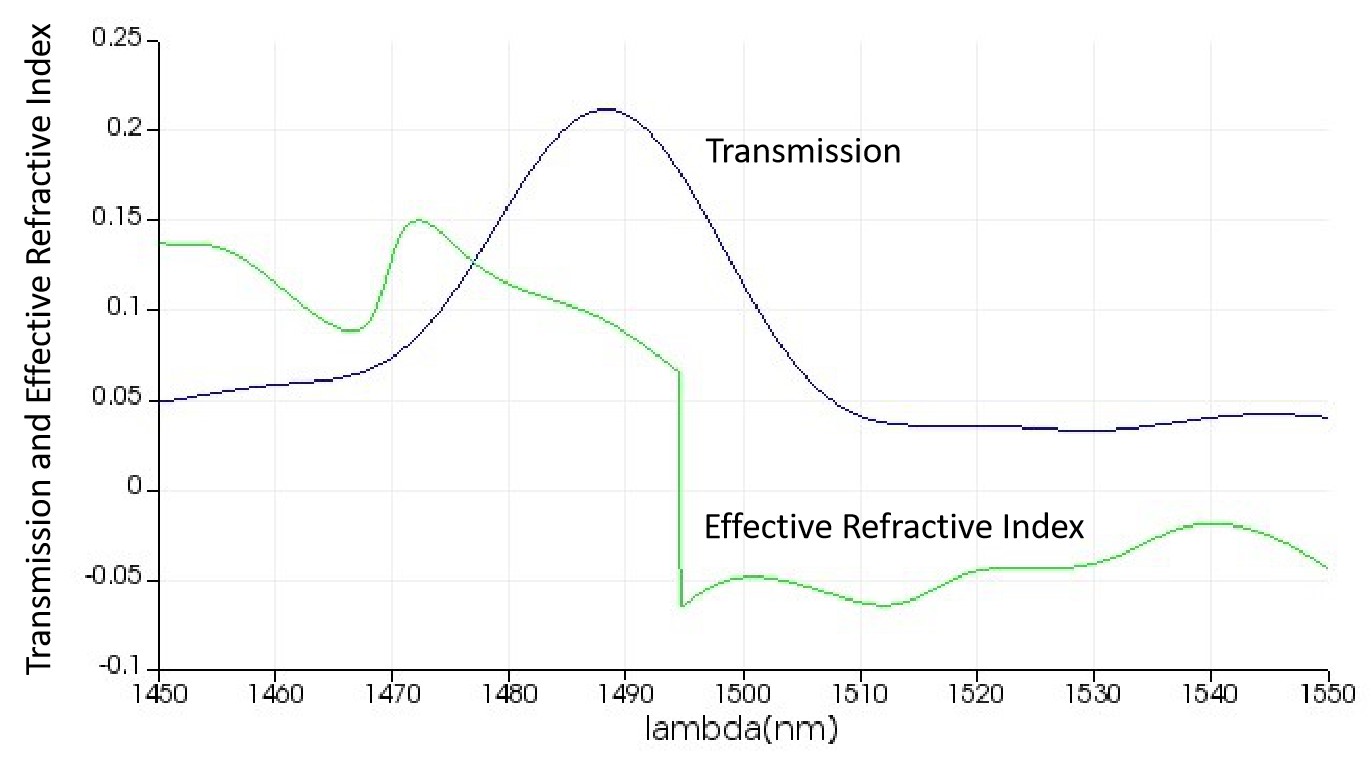}
{ Transmission and the value of the effective refractive index.\label{fig:transmission}}

We know that a non-trivial magnetic response is a necessary condition to achieve a refractive index equal to zero. The refractive index  \(n={n}'+{n}''=\sqrt{\mu \varepsilon }\) is generally complex-valued, determined by the complex frequency-dependent permittivity \(\varepsilon (\omega )\) and permeability \(\mu (\omega )\) of a material. In our case  \(\mu (\omega )\) equals zero at 1495 nm, which means both the real part of refractive index and imaginary index equal zero, which matches our simulation shown in Fig. \ref{fig:refractive_index}. Between 1495nm and 1563 nm is the band gap where real index is still around zero, but imaginary index increases which causes loss. At 1563 nm, permittivity \(\varepsilon (\omega )\)  equal zero and above 1563 nm, the refractive index becomes negative.

\par

Also, the impedance value is given by \(Z=\sqrt{\mu /\varepsilon }\). Since this is a uNZ metamaterial, it has low impedance. In our case \(\mu (\omega )\)  equals zero at 1495 nm, which means the impedance Z is around zero, which matches our simulation shown in Fig. \ref{fig:impedance}. Between 1495nm and 1563 nm is the band gap where Z increases. At 1563 nm, permittivity \(\varepsilon (\omega )\)  equals zero, so Z is very large. Above 1563 nm, the impedance decreases due to permittivity \(\varepsilon (\omega )\)  increasing.

\par

At the $\Gamma$-point, the Q factor of mode 2 has a value of Q = 1.8* $10^4$ (\ref{fig:Q_mode_2}), which is high. That is why there is BIC around 1495 nm. But around k = 0.09, there is still a higher zero Q factor of 3.7*$10^4$ at 1486 nm, which also keeps the BIC. That is why the highest output is at 1486 nm Fig. \ref{fig:transmission}.

\Figure[ht!](topskip=0pt, botskip=0pt, midskip=0pt)[width=0.45\textwidth]{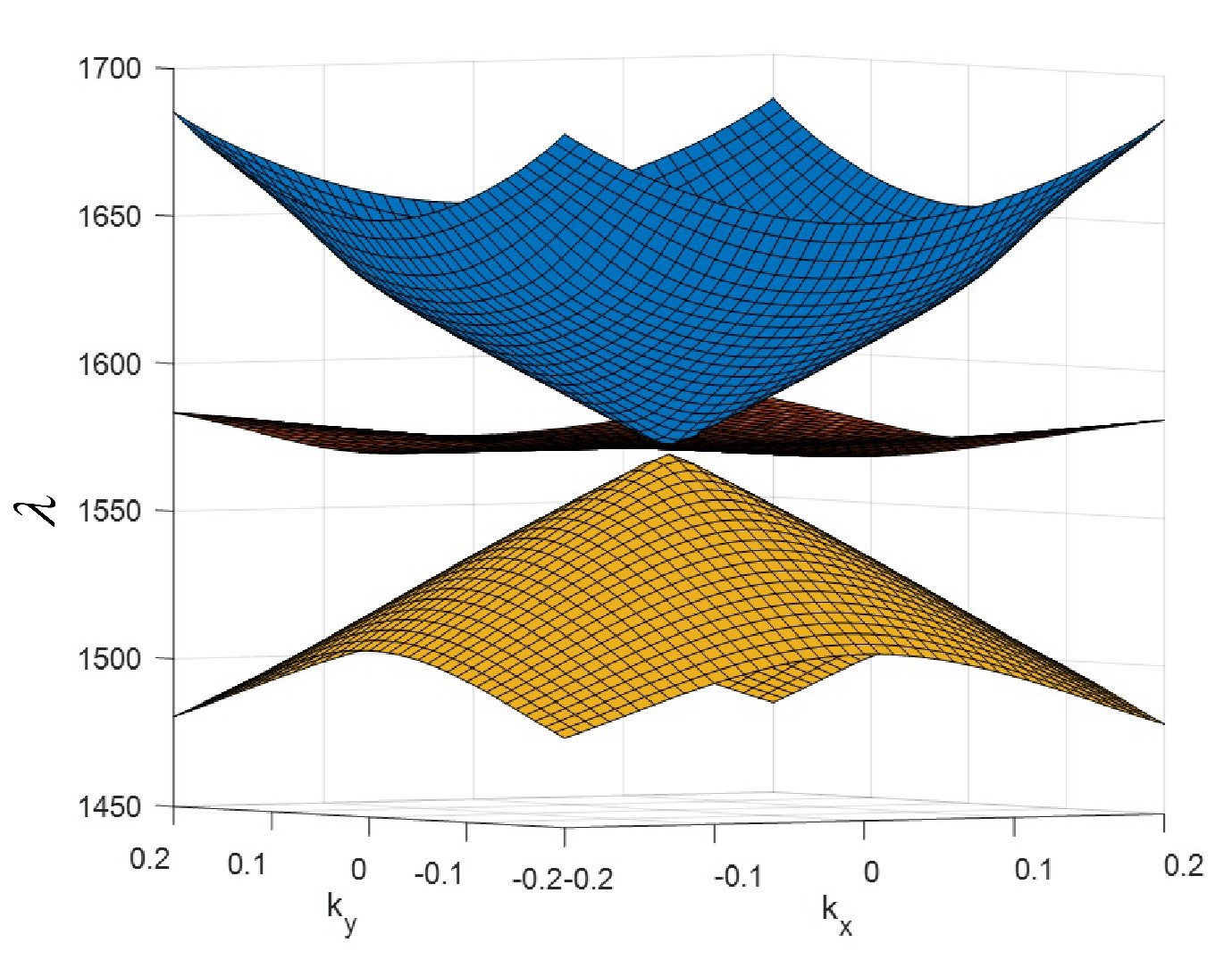}
{ Bound state in the continuum on eNZ metamaterials. The structure consists of a 3D Si pillar array. \label{fig:eNz_band}} 

\section{The BIC eNZ metamaterials}\label{sec:related}
Similarly, we can tune the Si parameters to achieve the BIC eNZ conditions. 


Fig. \ref{fig:eNz_band} shows that the Dirac cone is formed at the center of the Brillouin zone. Even at the gamma point, linear dispersion represents a finite impedance. The color represents the quality factor of the propagation mode, with a strong peak near the Dirac point. The monopole and dipole modes become degenerate and lossless simultaneously at an operating wavelength of 1570 nm. This behavior is predicted by CMT and confirmed by finite element analysis. As a result, the optimized metamaterial supports a zero-index mode at 1570 nm, which propagates without phase advance or radiative loss.

The BIC condition depends on the free-space wavevector, and therefore can only be satisfied for a single wavelength. As a result, the Q factor is finite and decreases as the operating frequency is detuned from zero index. The imaginary index signals the return of radiative loss for positive and negative index modes. The weak-coupling approximation is valid for resonators with high quality factor (here the Q factor is for radiation at z direction).
   	
Importantly, the zero-index mode is isotropic, and can be excited by incident waves from arbitrary directions. This is demonstrated by the band structure in the reduced Brillouin zone, which forms a Dirac cone at the $\Gamma$-point. Fig. \ref{fig:eNz_band} show the dispersion surfaces, and also indicates the Q factor at each point in k-space. The sharp cone at the operating wavelength indicates that the modes are simultaneously degenerate and lossless \cite{Maas_2013}.

\Figure[ht!](topskip=0pt, botskip=0pt, midskip=0pt)[width=0.45\textwidth]{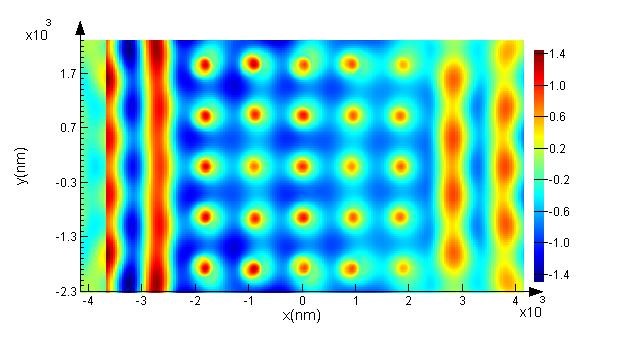}
{ BIC ZIM Profile of in-plane. The field profile shows the vertical component of the electric field through a horizontal cross-section of the array at $n = 0$. The color bar represents the intensity of electric field.\label{fig:field_xy_eNZ}}

\Figure[ht!](topskip=0pt, botskip=0pt, midskip=0pt)[width=0.45\textwidth]{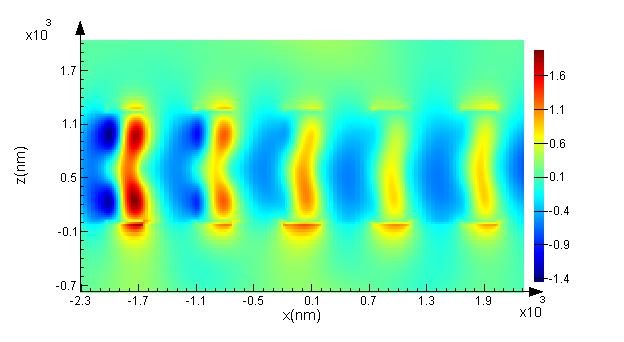}
{ BIC ZIM Profile of out-plane. The field profile shows the vertical component of the electric field through a vertical cross-section of the array at n = 0. The color bar represents the intensity of electric field.\label{fig:field_xz_eNZ}}

Fig. \ref{fig:field_xy_eNZ} shows in-plane field distribution of the BIC eNZ metamaterials. The field profile shows the vertical component of the electric field through a horizontal cross-section of the array at n = 0. The color bar represents the intensity of electric field.
 
Fig. \ref{fig:field_xz_eNZ} shows the filed distribution of the BIC eNZ metamaterials. The field profile shows the vertical component of the electric field through a vertical cross-section of the array at  $n = 0$. The color bar represents the intensity of electric field.


\Figure[ht!](topskip=0pt, botskip=0pt, midskip=0pt)[width=0.45\textwidth]{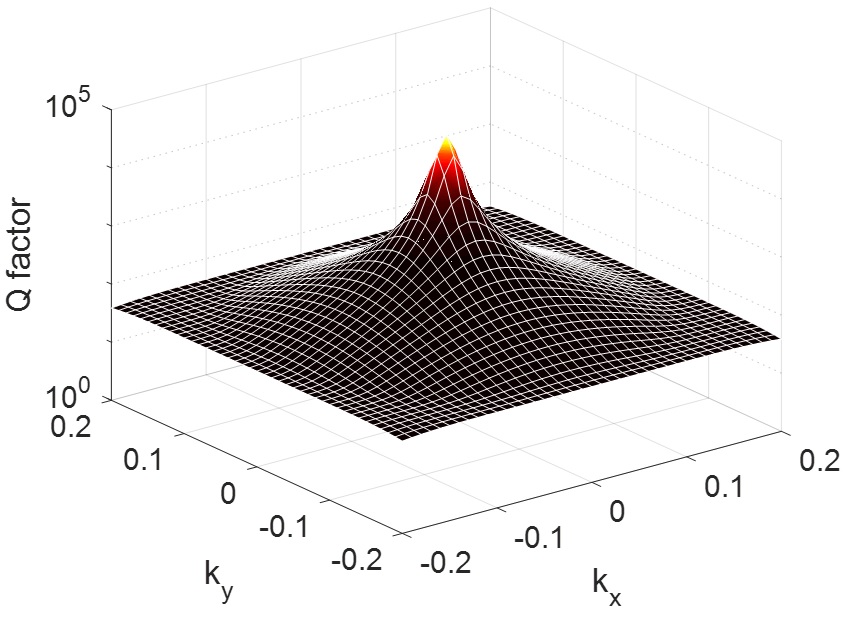}
{ The quality factor of propagation mode 1 is shown in color, which strongly peaks near the Dirac point.\label{fig:Q1_eNZ}}

\Figure[ht!](topskip=0pt, botskip=0pt, midskip=0pt)[width=0.45\textwidth]{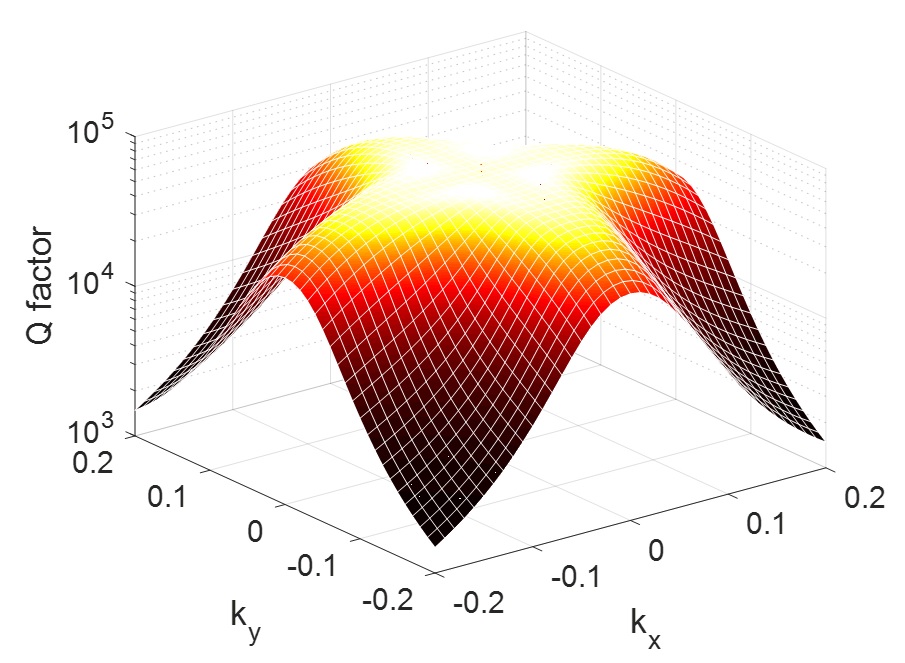}
{ The quality factor of propagation mode 2 is shown in color, which strongly peaks near the Dirac point.\label{fig:Q2_eNZ}}

\Figure[ht!](topskip=0pt, botskip=0pt, midskip=0pt)[width=0.45\textwidth]{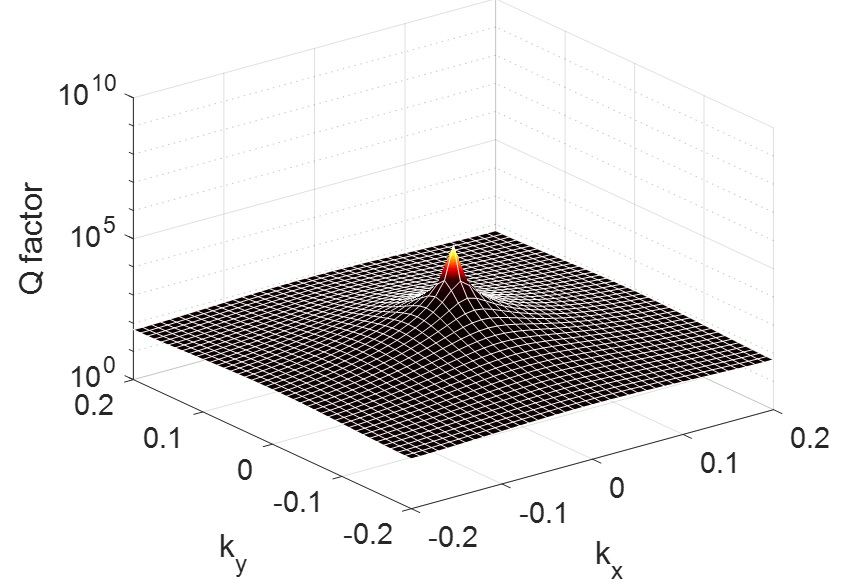}
{  The quality factor of propagation mode 3 is shown in color, which strongly peaks near the Dirac point.\label{fig:Q3_eNZ}}
     
  Near the center of the Brillouin zone a Dirac cone is formed indicating a finite impedance and zero index of refraction at the Dirac point. The quality factor is strongly peaked near the wavelength and is also isotropic, as shown in Figs. \ref{fig:Q1_eNZ}-\ref{fig:Q3_eNZ}. The optimized design achieves Q > $10^5$, up to two orders of magnitude higher than previous design \cite{Maas_2013}. Bound states at the Dirac-point propagate as lossless zero-index modes, which can exhibit exotic optical phenomena over large areas. These effects have so far been limited to hypothetical bulk media or infinitely tall metamaterial arrays. Because of the limited range of BIC ZIM, lossless propagation can now be implemented in planar devices and integrated with on-chip photonic devices. For example, we simulate a BIC ZIM plate irradiated by a planar Si waveguide. As in the case of 3D metamaterials, the input wave is coupled to the zero refractive index mode, which propagates through the slab without phase advance or loss of radiation.
 

\section{outlook}\label{sec:preliminaries}

 Structures with near-zero parameters first attracted the attention of researchers as pathological cases in the field of metamaterials.
Subsequent research discovered a number of unusual wave phenomena that challenged our understanding of light-matter interactions.
Examples presented here have shown that near-zero refractive index photonics exhibits very distinctive features in basic light-matter interaction processes including the propagation, scattering,
emission and confinement of light. Even more importantly, near zero
refractive index photonics also enables wave phenomena, such as tunnelling/supercoupling and geometry-invariant eigenmodes, which are exclusive to structures with near-zero parameters. At this point,
nonlinear optics, flexible photonics, quantum information processing and heat management seem to be some of the most promising areas. Ultimately, their technological impact will be determined by
our ability to develop high-quality and, in particular, low-loss structures with near-zero parameters. 

 \par
 
 BICs arise through several distinct mechanisms and exist in a wide range of material systems. Optical systems provide a clean and versatile platform for realizing different types of BICs because of nanofabrication technologies that enable the creation of customized photonic structures. An optical BIC exhibits an ultrahigh Q - which can increase the interaction time between light and matter by orders of magnitude \cite{Liao_2020_IEEE_Quantum}. In addition to the high-Q applications described above, there are many more opportunities in, for example, nonlinearity enhancement and quantum optical applications that have not been explored. The long-range interactions in Fabry-Perot BICs may be useful for nanophotonic circuits\cite{SatoY} and for quantum information processing. It has also been proposed that the light intensity may act as a tuning parameter in nonlinear materials, which may enable robust BICs\cite{Bulgakov}, tunable channel dropping, light storage and release, and frequency comb generation\cite{Pichugin}. Finally, it was shown that particle statistics can be used to modify properties of BICs\cite{Crespi}.  Considering the many types of BICs, a natural question is whether a common concept underlies them all other than the vanishing of coupling to radiation through interference.   Bound states at the uNZ/eNZ as lossless zero-index modes can exhibit exotic optical phenomena over large areas. We explore bound state in the continuum with zero refractive index in uNZ/eNZ metamaterials. By adjusting the unit-cell dimensions including radius, pitch and height, we can completely eliminate the out-of-plane radiation. There are many more opportunities in, for example, nonlinearity enhancement and quantum optical applications that have not been explored. To this end, the topological interpretation of BIC seems promising. The topological arguments may guide the discovery of BIC and new ways to trap waves, which may also exist in quasi-particle systems such as magnons, polaritons, polarons and any one. Because BIC defy conventional wisdom and provide new ways to confine waves, their realization in different material systems are certain to provide even more surprises and advances in both fundamental physics and technological applications such as integrated photonics components and lasers \cite{Peng_2019}, \cite{Zeng_2019}, \cite{Peng_2019_NEMO}.
 
 \section{Conclusion}
 In this paper, we have reviewed the BIC and ZIM metamaterials, including fundamental principle, the inter-relation between the BIC and ZIM phenomena, realization structures and potential applications. It has been shown that the multi-mode interference is the underlying physics of the exotic BIC and ZIM structures.  In particular, the Si uNZ and eNZ metamaterials have been theoretically studied, simulated and designed. Theoretical formulas have been shown to design the Si uNZ and eNZ structures. The review can provide guidelines to design the BIC and ZIM structures for many potential important applications such as the integrated Si photonics and novel BIC and ZIM lasers.

 \EOD

\end{document}